\documentclass[conference]{IEEEtran}

\usepackage[utf8]{inputenc}
\usepackage[T1]{fontenc}
\usepackage{tabularx}
\usepackage{placeins}
\usepackage{algorithm2e}
\usepackage{amsmath}
\usepackage{graphicx}
\usepackage{comment}
\usepackage[breaklinks=true]{hyperref}
\usepackage{listings}

\usepackage{multirow}
\usepackage{breakurl}

\usepackage{float} 

\includecomment{comment} 
\DeclareUnicodeCharacter{2264}{<=}
\begin{document}

\title{\textbf{Discovering Encrypted Bot and Ransomware Payloads Through Memory Inspection Without A Priori Knowledge}}

\author{\IEEEauthorblockN{Peter McLaren, William J Buchanan, Gordon Russell, Zhiyuan Tan}
\IEEEauthorblockA{School of Computing, Edinburgh Napier University, Edinburgh, UK.}
}

\maketitle

\begin{abstract}
Malware writers frequently try to hide the activities of their agents within tunnelled traffic. Within the Kill Chain model the infection time is often measured in seconds, and if the infection is not detected and blocked, the malware agent, such as a bot, will often then set up a secret channel to communicate with its controller. In the case of ransomware the communicated payload may include the encryption key used for the infected host to register its infection. As a malware infection can spread across a network in seconds, it is often important to detect its activities \emph{on the air}, \emph{in memory} and \emph{at-rest}. Malware increasingly uses encrypted channels for communicating with their controllers. This paper presents a new approach to discovering the cryptographic artefacts of real malware clients that use cryptographic libraries of the Microsoft Windows operating system. This enables malware secret communications to be discovered without any prior malware knowledge.
\end{abstract}


\section{Introduction}

Malicious actors employ secure channels to hide communications from detection agents. These channels enable activities such as the installation of exploit kits, distribution of malware and adware, and communicating useful information to controllers. So, knowledge of channel contents is unknown while malicious use of secure channel surges - a cloud vendor blocked 1.7 billion threats using TLS in he second half of 2018 \cite{Desai2019}. Malware classes that frequently uses TLS secure channel for communications are bots and ransomware.

The ransomware element of an attack can be defined as the weaponisation part, where a ransomware element can be packaged with a defined distribution and infection methods, and then targeted as required. The first phase of ransomware, such as WannaCry, was often fairly scattergun in its targets, but the usage of ransomware in targeted attacks increases. In 2019, Symantec found that enterprise-targeting ransomware showed a yearly increase of 12\% and attacks on mobile devices rose by 33\% \cite{Symantec}. With WannaCry we saw the weaponisation of ransomware, which used the Eternal Blue vulnerability for its infection. The NHS in the UK was but one organization of many who suffered major outages from the ransomware infection \cite{hoeksma2017nhs}.

Modern bot malware is generally multi-purpose. Once installed on a client, an external controller determines bot activities through issued commands. Encrypted controller-to-bot channels prevent defenders from discovering commands. Ransomware clients are generally single purpose in that users pay to recover document or device access. Crypto-ransomware, where documents on an infected client are encrypted and a payment is required, typically in Bitcoins, for the decryption key, is a common variant \cite{kharraz2015cutting}. Communications between crypto-ransomware clients and controllers may include useful information such as encryption keys.
 
Current methods of dealing with malicious use of secure channels have limitations. Whereas in unencrypted channels, payload inspection provides knowledge of malware activity, with encrypted channels, detection methods rely on discovering anomalies between benign and malicious activity. These methods assist in the detection and possible prevention of possible malicious activity but cannot provide detailed knowledge of the malicious activity.

This paper investigates decrypting TLS communications of real-world malware. A framework uses a standard approach for decrypting TLS traffic to analyse and decrypt the secure communications. For malware, performance challenges can result from malware use of different cryptographic libraries. So, the framework is extended to accommodate the Windows cryptographic library. Experiments evaluate decrypting real bot and ransomware command and control communications using the extension. The contribution is a novel method to discovering cryptographic artefacts used by real malware. As these are discovered in single memory extracts and decrypted in less than a second, the communicated activities of unknown malware can be discovered.

The rest of the paper is structured as follows. Related research on malware command and control channels is presented in Section II. Section III discusses sourcing of real-world malware samples while Section IV evaluates and discusses the limitations of decrypting using a standard TLS decryption methodology. An new approach for decrypting traffic using Windows cryptographic libraries is presented in Section V. The results are presented and discussed in Section VI and conclusions drawn in Section VII.


\section{Related Work}
A number of papers focus on memory inspection to discover the malware threat using calls to APIs. Gupta et al \cite{gupta2016malware} analysed API calls within Windows and mapped a total of 534 important API calls with 26 categories (A-Z). These were then used to identify five types of malware (Worm, Trojan-Downloader, Trojan-Spy, Trojan-Dropper and Backdoor). Hampton et al \cite{hampton2018ransomware} furthered this work by analysing 14 strains of ransomware on Windows platforms and created mappings of the API call frequencies.

Other studies detect the presence of malware command and control channels. Signature-based detection systems that check for known byte sequences in packet headers or payloads may be less successful with new malware variants or encrypted traffic. So, anomaly detection methods using data mining methods \cite{shekhawat2019feature} can distinguish benign from malware traffic. For example, features such as correlations between malware channel request and response times \cite{gu2008botsniffer}, short packet sizes \cite{ma2010novel}, TLS header information \cite{anderson2016deciphering}, or a combination of features \cite{mclaren2017mining} are possible differentiators. The features facilitate malware detection and prevention but the encrypted contents are hidden.

Controller emulation can discover malware client plaintext. Logging client requests may provide useful insights although cited challenges may be environment security, scalability for large botnets, and transparency, where malware detects the presence of a test environment and terminates \cite{lin2013botnet}. Although a controlled environment is advantageous
\cite{lee2009framework} and can be used to detect adversaral activities \cite{sentanoe2017virtual} emulator drawbacks for decrypting real-world malware communications may be not knowing valid controller responses but, perhaps more importantly, the malware must be known a priori to execute in the environment.

Patil et al \cite{patil2019windows} define an investigation framework for analysing captured memory for the detection of malware using information from processes, running threads, opened registry keys, and user authentication details. Google, too, focus on memory inspection for the detection of malware \cite{thioux2019system} and use a number of virtual machines to detect anomalous behaviour.

Feichtner et al \cite{feichtner2018automated} defines a method of detecting cryptographic misuse in Apple iOS applications. Their work uses a decompilation method to analyse the code calls to the core cryptographic libraries. In their analysis they found that 82\% of the applications sampled had a cryptographic flaw. For this they defined six main rules to identify a cryptographic flaw. These included: the usage of the ECB mode for encryption; the usage of a non-random IV for CBC encryption; and the usage of constant encryption keys. Most of flaws found related to the usage of a non-random Initialization Vector (IV) and the usage of constant encryption keys. They also found that 27\% of the sampled apps used ECB (Electronic Code Book) for the symmetric key encryption.

A priori malware knowledge may also enable plaintext discovery. While virtual machine environments, such as DRAKVUF \cite{Lengyel2014}, support malware dynamic analysis, such solutions succeed only where actual, or suspected, malware is obtained.  Our approach requires no prior knowledge so discovering the plaintext in the encrypted communications of unknown malware is possible.

Discovery of malware cryptographic artefacts may enable their communications to be decrypted. For example, researchers have discovered TLS encryption keys in memory \cite{taubmann2016tlskex}. Initialization vectors were not discovered, which for commonly-used AES-GCM encryption is necessary, so plaintext was not derived. Furthermore, whereas keys were discovered in Linux memory, malware is still predominantly Windows-focused \cite{anderson2016deciphering}.

\section{Sourcing Malware Samples}
Malware communications analysis ideally executes real malware samples. Samples of recent provenance are preferable as these may reflect the approaches of modern malware authors, Maintained online databases provide guides to current usage. For SSL traffic, the SSL Blacklist website \cite{SSLBlacklist} lists bot and ransomware clients in reverse chronological order of awareness. From the list, 35 potential malware entities were identified. For each, multiple client executable samples were downloaded in compressed, password protected from the VirusShare website \cite{VirusShare}. 
The compressed files are securely copied to the client. on which Windows Defender is disabled, the compressed files uncompressed and executed. As the responder acting as a malware controller was not configured to respond appropriately, only three executables successfully established a TLS connection including a data exchange with a server running the OpenSSL server application \cite{OpenSSL}.
The three malware executables are listed in Table \ref{tab:MalwareSamples} followed by pertinent background.

\begin{table}[h]
    \centering
    \caption{Malware Samples}
    \begin{tabular}{l|l|l}
    \hline
    \textbf{Type}&\textbf{Class}&\textbf{MD5}\\
    \hline
    Bot&Zbot&eeef1e062c8011cabb23b3c833ff766a\\
    Ransomware&Torrentlocker&aeb5bb78ab442bc94bb94d968754e523\\
    Bot&Gozi&67a775879d3664456cb6a5026c518ca0\\
    \end{tabular}
    \label{tab:MalwareSamples}
\end{table}

\begin{figure}[!htb]
        \center{\includegraphics[width=0.4\textwidth]
        {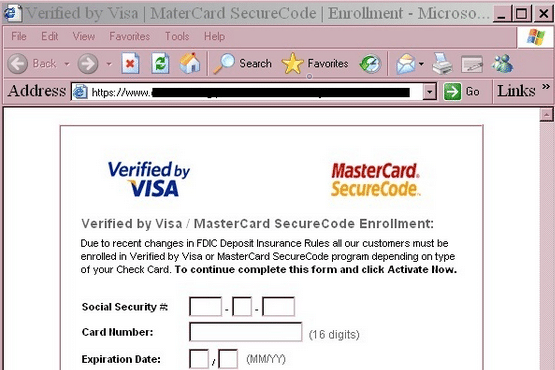}}
        \caption{\label{fig:ZbotFake}Zbot Fake \cite{Raju2016}}
\end{figure}

Zbot, also known as ZeuS or Zeus, is a well-known bot malware instance. Detected in 2006 \cite{TrendMicro2015}, Zbot is primarily known for stealing banking passwords by injecting code into a user browser as illustrated in Figure \ref{fig:ZbotFake}. Other Zbot functionality includes extracting information such as browser history and cookies, certificates, and mail account information as well as perform actions such as manipulate local files, install ransomware, log keystrokes, take screenshots, and manage a botnet of other infected computers \cite{ZbotSymantec} \cite{Panda2017}. Although Zbot previously used the HTTP protocol for communications, information and commands are now generally concealed with TLS.

\begin{figure*}[!htb]
        \center{\includegraphics[width=0.6\textwidth]
        {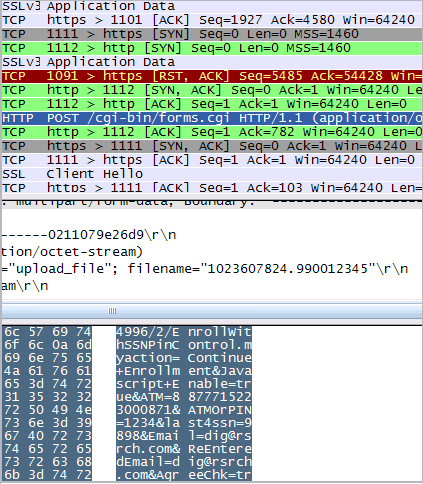}}
        \caption{\label{fig:GoziTheft}Gozi Data Theft \cite{Jackson2007}}
\end{figure*}

Gozi, also known as Ursnif inter alia, is also an information-stealing bot. Detected in 2007 \cite{Jackson2007}, Gozi is commonly used by malicious actors for stealing and other confidential banking information \cite{Alvarez2018} \cite{Brumaghin2018} \cite{Garnaeva2016} as shown in \ref{fig:GoziTheft}. Although functions include theft of cookies and email credential, and logging of keystrokes and browsing activity \cite{Mohanta2018}, a key feature is intercepting network traffic to hijack financial transactions. For example, when a money transfer is detected, Gozi issues an encrypted message through its command and control server to prevent the correct transfer and redirect the funds to a controlled account \cite{Weller2016}. 

\begin{figure}[!htb]
        \center{\includegraphics[width=0.5\textwidth]
        {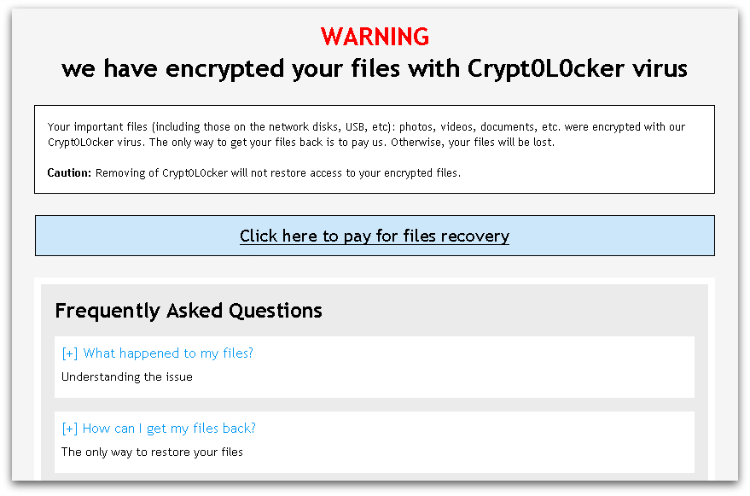}}
        \caption{\label{fig:TorrentLockerRansom}TorrentLocker/Crypt0L0cker Warning}
\end{figure}

TorrentLocker is an instance of crypto-ransomware. Known since 2014, sufficiently similar to CryptoLocker to also be known as Crypt0L0cker, it encrypts user documents including pictures, advises the user of its action, and demands payment using Bitcoins \cite{M.Leveille2014} as indicated in Figure \ref{fig:TorrentLockerRansom}. Although client-controller communications were previously encrypted using XOR, TLS is now a common communications mechanism \cite{M.Leveille2016}. Information transmitted to the controller includes the ransom page, the encryption key which is RSA-encrypted with a TorrentLocker public key, counts of encrypted files, address book contacts, email credentials, and logs \cite{M.Leveille2014}.

\section{Approach}
The analysis framework executes in a virtualized environment, A hypervisor supports a virtual machine monitor executing on a privileged virtual machine, a suspect client virtual machine executing malware samples, and virtual machines providing server functionality for client communications. The framework extracts read/write client virtual machine memory, analyses memory extracts to discover small sets of candidate cryptographic artefacts, and decrypts  encrypted network traffic until a decrypt is validated. The framework accommodates SSH and TLS protocols and encryption algorithms such as AES and ChaCha20.

\begin{figure*}[!htb]
        \center{\includegraphics[width=0.9\textwidth]
        {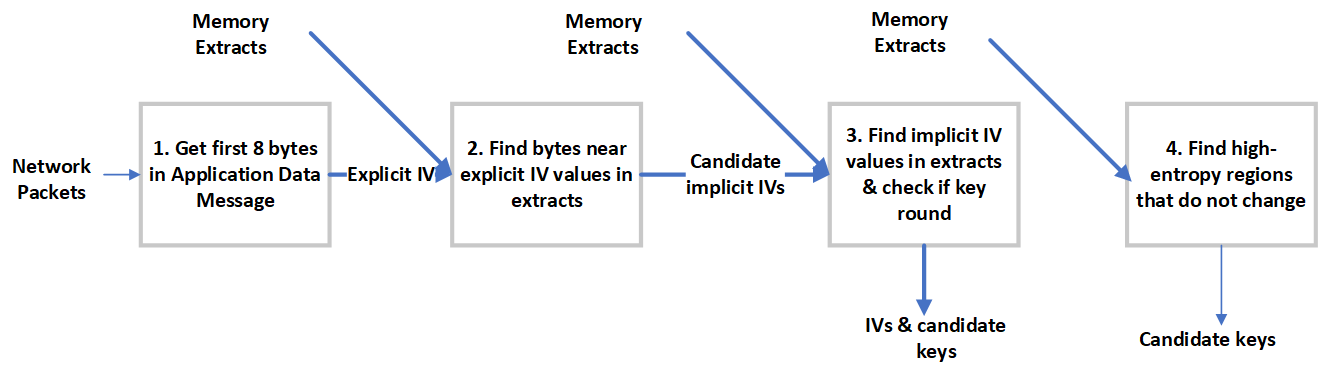}}
        \caption{\label{fig:GCMMemoryAnalysis}GCM Memory Analysis Approach}
\end{figure*}

A standard TLS extension searches memory extracts for key blocks that a pseudo-random number generator (PRNG) create after the handshake. For AES-GCM, the agreed encryption algorithm for each malware sample, key blocks contain client and server encryption keys, and client and server implicit initialization vector (IV) segments. MemDecrypt memory analysis searches for key blocks following the process illustrated in \ref{fig:GCMMemoryAnalysis} where the explicit IV segment in an Application Data network packet enables searches for candidate implicit IVs, which enable candidate key block discovery. Candidate implicit IVs are memory extract segments co-located with explicit IV segment values. Candidate key blocks are memory extract segments co-located with candidate implicit IV segments and where the key block client key and server key exceed a threshold. 

\subsection{Test Environment}
The Xen Project 4.4.1 hypervisor runs on a Core 2 Duo Dell personal computer with 40\,GB of disk storage and 3\,GB of RAM,  It supports a privileged hypervisor console running Debian release 3.16.0-4-amd64 version and the MemDecrypt framework. and three unprivileged virtual machines. Experiments execute on a Windows client and a Linux server virtual machine. The client runs the Windows 10 (10.0.16299) operating system with 2 GB of memory and 40\,GB of disk, and the  server runs an Ubuntu 14.04 build (“Trusty”) with 512\,MB of allocated memory and 4\,GB of disk storage. 

The environment is configured for malware containment. The malware client is prevented from communicating with external servers to prevent possible corruption of other environments. So, an additional Linux machine running Ubuntu 14.04 build (“Trusty”) with 512\,MB of allocated memory and 4\,GB of disk storage is established as a DNS server using the ‘dnsmasq’ package. Responses to benign DNS requests, such as *.microsoft.com, return the DNS server IP address and to other requests the IP address of the target TLS server. For the first experiment, debug mode was enabled to log keys, IVs and plaintext. The OpenSSL server command used was:\\

\textit{openssl s\_server -accept 443 -debug -cert crt.pem -key key.pem -WWW}

\subsection{Results}
With Zbot, the application of the standard TLS extension to memory extraction and analysis, the decrypt analysis component  was projected to require approximately 34 hours to identify correct artefacts. By excluding artefacts less than 1000 bytes apart in memory extracts, this reduced to 15 minutes so the combined analysis duration was 38 minutes. Although quicker than brute-force, cryptanalytic, and side-channel approaches, the duration may be insufficient for practical application in live scenarios. Furthermore, the duration is substantially longer than earlier experiments with, for example, the OpenSSL library, warranting further analysis. 

\begin{figure*}[!htb]
        \center{\includegraphics[width=1.0\textwidth]
        {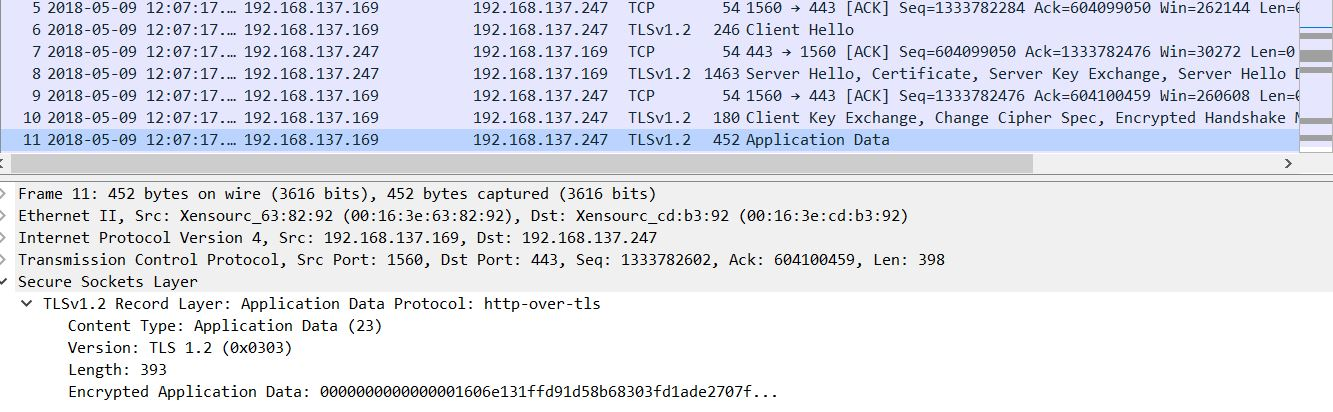}}
        \caption{\label{fig:GoziHandshake}Gozi Handshake \& Application Data message}
\end{figure*}

Two factors cause this increase. One is the 8-byte explicit IV segment obtained from an Application Data Message. For Zbot, the first explicit IV segment is \emph{'0x0000000000000001'} as illustrated in the highlighted section of the Wireshark packet capture in Figure \ref{fig:GoziHandshake}. This byte sequence occurs more frequently than randomly generated explicit IVs in memory extracts. So, when the explicit IV is used to discover possible four-byte implicit IV segments, 578,629 possible instances were found. Entropy measure thresholds reduced the candidate key block size to  23,361. By contrast, an experiment with an OpenSSL client application yielded three candidate implicit IVs and 79 candidate keys. 

The other factor is masquerading. To evade detection, malware applications may camouflage their activities and one such mechanism is masquerading as a benign application so the malware may also be known as a ‘trojan’. In the Windows environment, examples of benign applications used for masquerading includes the Edge browser and Windows Explorer. However, when Zbot masquerades as the Windows Explorer, for example, the data collection component extracts 265 read/write memory files totalling 73.2\,MB for each separate extraction. The combination of these two factors leads to large sets of candidate keys and IVs. 

\section{Windows Library Extension}
Memory extract features suggest a more efficient alternative. Using the session key and IV from OpenSSL server logs, a search of malware client application memory yielded interesting facts: the key occurs frequently in different extract files, memory extract files sizes containing the key are within specific ranges, and two unusual ASCII strings are present near encryption key locations in the memory extract files.

The repeated occurrence of the key in memory extracts may result from data protection or an absence of data cleansing. After a TLS handshake when client and server keys have been generated by a pseudo-random generator, the keys may be copied to record data structures for simplified access by the encryption process. The malware may copy the keys repeatedly to ensure access. Alternatively, the malware writer may copy keys on different occasions but fail to cleanse the source or copy. In any case, this feature is not used in the extension.

\begin{figure*}[!htb]
        \center{\includegraphics[width=0.7\textwidth]
        {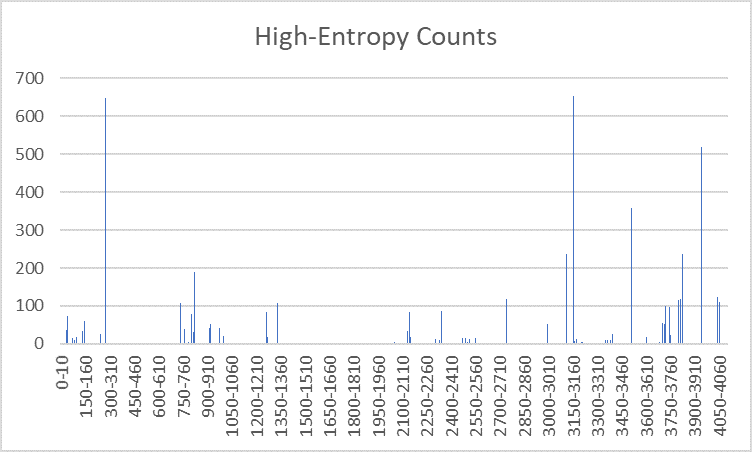}}
        \caption{\label{fig:HighEntropy}Windows Explorer High-entropy Regions}
\end{figure*}

Sizes of memory extract files containing encryption keys ranged between 2 MB and 4 MB. Consistent with prior SSH and TLS investigations, the size probably originates from an application memory allocation request (‘malloc’) for a data structure to hold the encryption, or decryption fields such as keys, encrypt/decrypt flags, key length, mode, etc. As illustrated in Figure \ref{fig:HighEntropy} which maps the number of segments above a 4.5 threshold (Y-axis) against the extracted size (X-axis), the distribution of high-entropy counts in malware application memory suggests that prioritizing regions for analysis may speed up the IV and key discovery process.

The presence of specific ASCII strings in memory extracts containing keys is more significant. The strings are \emph{‘3LLS’} and \emph{‘KSSM’}, or in big-endian format" \emph{'SSL3’} and \emph{‘MSSK’}. Researchers identified ‘MSSK’ in the Windows security policy application, Local Security Authority Subsystem Service (LSASS) as a possible acronym for ‘Microsoft Symmetric Key’ or ‘Microsoft Symmetric Key’ \cite{Kambic2016}. ‘SSL3’ may refer to the deprecated forerunner of TLS, SSLv3. Kambic identified probable fields in the undocumented LSASS data structure including: encryption data structure sizes; TLS version; and the encryption key. The field identified by Kambic as the probable IV field is inconsistent with MemDecrypt memory extracts. The implicit IV is located approximately 20 bytes after the ‘3LLS’ string, and the key approximately 30 bytes after the ‘KSSM’ string. Although random occurrences are possible, the strings provide good indicators for identifying candidate memory extracts containing cryptographic artefacts when Microsoft security libraries are used.

A MemDecrypt extension to decrypt TLS communications from executables that use Microsoft security libraries accommodates these features. Microsoft encryption libraries are assumed When TLS Application Data messages contain explicit IV values of ‘x0000000000000001’, Additional techniques such as the identification of executable linked libraries might validate this assumption. Extract file sizes are banded to prioritise medium-sized files. The extract files are searched for the ASCII strings and fields in near locations in the same extracts and of sufficient entropy to be candidate keys and IVs are identified. The Microsoft memory analysis algorithm is shown in Algorithm \ref{alg:MicrosoftLibraryDecrypt}. The banding is wider, and the maximum allowable distances in memory between '3LLS' and a candidate IV, and 'KSSM' and a candidate key exceed the empirically observed values to allow for potential data structure changes, as may result from operating system upgrades. The entropy thresholds for 'IVsize' and 'keysize' are set to 1.5 and 4.5 respectively based on previous experiments with PRNG functions. If the Microsoft library extension fails to find cryptographic artefacts, the TLS extension provides a fall-back.\\
 
\begin{algorithm}
\SetAlgoLined
\KwData{Extracts folder, entropy thresholds, keysize}
\KwResult{Z = candidate keys, Y = candidate IVs}
\For{file in folder}{
    \If{1 MB < size(file) < 8MB }{
        Band1 += file\;
    }
    \If{0 MB < size(file) < 1MB }{
        Band2 += file\;
    }  
    \If{8 MB < size(file)  }{
        Band3 += file\;
    }
}
IVsize = 4\;
\For {each file in Bands 1-3}{
     \If {‘KSSM’ in extract} {
        start = location (‘3LLS’)\;
        \For {i = start to MaxIVdistance inc by 4}{
            s = extract[i:i+4]\;
            \If {entropy(s) > threshold(IVsize)}{
                Y += s
                }
        }
        Start = location ‘KSSM’\;
        \For {i = start to MaxKeydistance increment by 4}{
            s = extract[i:i+keysize]\;
            \If {entropy (s) > threshold(keysize)}{
                Z += s\;
            }
        }
    }
}
\caption{Windows Library Memory Analysis}
\label{alg:MicrosoftLibraryDecrypt}
\end{algorithm}

\section{Windows Library Extension Evaluation}  
\begin{figure*}[!htb]
        \center{\includegraphics[width=0.8\textwidth]
        {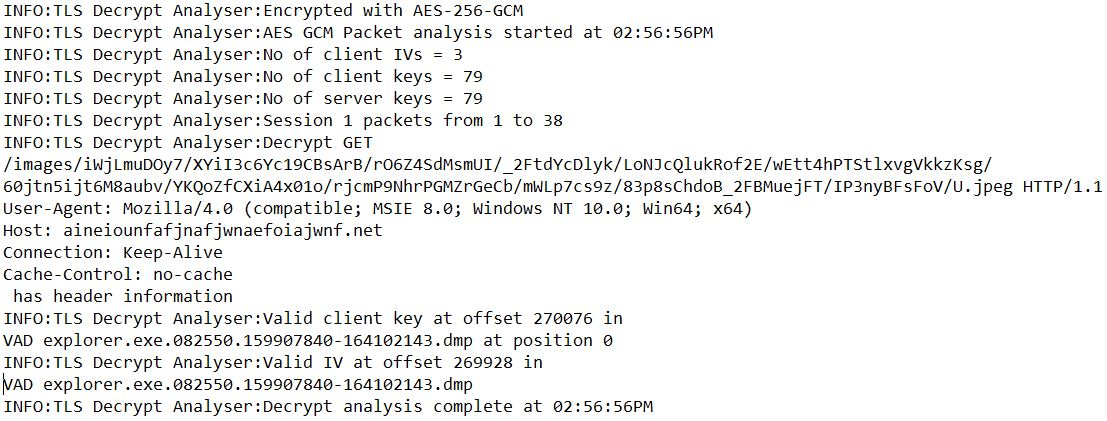}}
        \caption{\label{fig:ZbotDecrypt}Zbot Decrypt Log}
\end{figure*}

\begin{figure*}[!htb]
        \center{\includegraphics[width=0.65\textwidth]
        {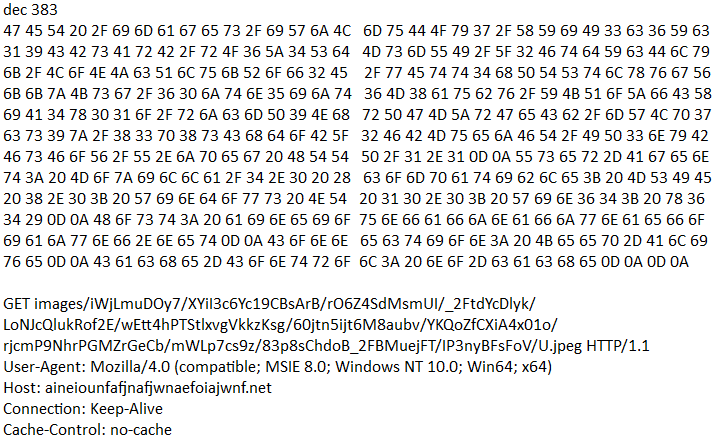}}
        \caption{\label{fig:ServerHexAscii}Zbot Server Log}
\end{figure*}

Experiments were executed to evaluate the Windows library extension. Each malware sample was executed on a Windows 10 client, memory extracted and Ubuntu OpenSSL server logs collected. Analysis decrypts were validated by evaluating compliance with HTTP 1.1 and comparison with server logs. An example of Zbot decrypted analysis output is illustrated in Figure \ref{fig:ZbotDecrypt} and verification provided by the OpenSSL server log shown in Figure \ref{fig:ServerHexAscii}. Each Zbot, Gozi, and TorrentLocker samples decrypted with 100\% success. Decrypt output examples for all malware samples are shown in Table \ref{tab:MalwareDecryptExamples}. Host names and GET image names vary for different test runs, and furthermore, Gozi decrypts produce POST as well as GET requests.

\begin{table}[h]
    \centering
    \caption{Malware Decrypt Analysis Output Examples}
    \begin{tabularx}{\textwidth}{c|p{6cm}}\cline{1-2}
    \textbf{Malware}&\textbf{Decrypt}\\\cline{1-2}
    \multirow{2}{6em}{Zbot} &GET /images/iWjLmuDOy7/XYiI3c6Yc19CBsArB \\ 
                            &/rO6Z4SdMsmUI/\_2FtdYcDlyk/LoNJcQlukRof2E/\\ 
                            &wEtt4hPTStlxvgVkkzKsg/60jtn5ijtM8aubv/YKQoZ\\
                            &fCXiA4x01o/rjcmP9NhrPGMZrGeb/mWLp7cs9z/83p8s\\
                            &ChdoB\_2FBMuejFT/IP3nyBFsFoV/U.jpeg HTTP/1.1\\
                            &User-Agent: Mozilla/4.0 (compatible; MSIE 8.0;\\
                            &Windows NT 10.0; Win64; x64)\\
                            &Host: aineiounfafjnafjwnaefoiajwnf.net\\
                            &Connection: Keep-Alive\\
                            &Cache-Control: no-cache\\\cline{1-2}
  \multirow{2}{6em}{TorrentLocker}&POST /topic.php HTTP/1.1 \\    
  &Accept: */*\\
  &Host: orbfoz.drinkmilks.org \\
  &Content-Length: 176 \\
  &Cache-Control: no-cache\\\cline{1-2}
    \multirow{2}{6em}{Gozi} &GET/images/SZK2b21jT4lHGlPQ/dzUZdQ7R6GQ9vDq/\\
                            &\_2Bubb8ji761TySLT2/0hh3OGhrh/Gp933q\_2\\
                            &BuLnl3Zz6zem/PlUvBGkYtrVJ8DUn8vN/LcmmaA101Yc\\
                            &rEmAU5RwlEy/zK8Z9xa06n3R/g0EFUOhv/8sDfI2Goa38\\
                            &03Voj6biFmRG/JDejp7Gffn/y7.jpeg HTTP/1.1\\
                            &User-Agent: Mozilla/4.0 (compatible; MSIE 8.0; \\
                            &Windows NT 10.0; Win64; x64)\\
                            &Host: wjenqwdqwdwdqwd.com\\
                            &Connection: Keep-Alive\\
                            &Cache-Control: no-cache\\\cline{1-2}
    \end{tabularx}
    \label{tab:MalwareDecryptExamples}
\end{table}

Analysis component durations for each malware sample confirm the extension’s performance. As illustrated in Table \ref{tab:MalwareTimes} the maximum combined duration for memory analysis and decrypt analysis is below 1 second, a direct consequence of reduced candidate cryptographic artefact set sizes. With the Microsoft library extension, the set sizes range between three and six, and IV set sizes between 79 and 483.

\begin{table}[h]
    \centering
    \caption{Malware Extension Analysis Times (secs)}
    \begin{tabular}{l|c|c}
    \hline
    \textbf{}&\textbf{Memory Analysis}&\textbf{Decrypt Analysis}\\
    \hline
    Maximum&0.21&0.41\\
    Minimum&0.18&0.03\\
    Mean&0.17&0.16\\
    Standard Deviation&0.03&0.18\\    
    \hline
    \end{tabular}
    \label{tab:MalwareTimes}
\end{table}
\subsection{Analysis}
The small experimental set size might inhibit complete confidence in the extension's capacity to decrypt encrypted malware command and control traffic. When malware writers have developed custom security routines, analysts have broken them easily broken so known cryptographic libraries are more commonly used. Microsoft library presents a good opportunity for malware writers being pre-loaded with a Windows operating system. Use of libraries such as OpenSSL would require additional download increasing the risk of detection. It is concluded that payloads of secure TLS communications between malware clients executing on Windows clients and their controllers can be rapidly discovered.

Decrypting a single request is not conclusive. This outcome is determined by testbed configuration as the server responds with OK to any TLS requests, In the absence of a reasonable response, the client may terminate or cease further communications with the controller. Also, decryption may not necessarily provide useable information, particularly where the plaintext includes a secondary encryption layer as with TorrentLocker. However, having discovered the key and IV, a complete session is decryptable.

\section{Conclusions}
Rapid decryption of live TLS malware traffic offers exciting prospects. For instance, by permitting the client to communicate with its controller in a managed environment, knowledge such as client-controller interaction details may contribute to enhanced malware defences. Furthermore, decrypting unmanaged communications between malware and controller may provide ransomware keys or stolen banking details. Future work will explore these opportunities as well as expanding the range of tested malware clients.

\bibliographystyle{IEEEtran}
\bibliography{main}

\end{document}